%% file: mKfollowup.tex
\documentclass[
twocolumn,
]{revtex4-2}
\usepackage{amsmath}
\usepackage[draft]{graphicx}
\usepackage{xcolor}
\usepackage[colorlinks]{hyperref}
\usepackage{siunitx} 
\usepackage[normalem]{ulem}

\begin{document}
	\title{Rydberg excitons in Cu$_2$O at millikelvin temperatures}
	\date{\today}
	\author{J.~Heck\"otter}
	\affiliation{Experimentelle Physik 2, Technische Universit\"at Dortmund, 44221 Dortmund, Germany}
		\author{D.~Janas}
	\affiliation{Experimentelle Physik 2, Technische Universit\"at Dortmund, 44221 Dortmund, Germany}
		\author{M.~Aßmann}
	\affiliation{Experimentelle Physik 2, Technische Universit\"at Dortmund, 44221 Dortmund, Germany}
		\author{M.~Bayer}
	\affiliation{Experimentelle Physik 2, Technische Universit\"at Dortmund, 44221 Dortmund, Germany}

	\begin{abstract}
		Rydberg excitons in the semiconductor Cu$_2$O have been observed in absorption experiments up to a principal quantum number of $n=28$ at millikelvin temperatures~\cite{heckotterExperimentalLimitationExtending2020}. 
		Here, we extend the experimental parameter space by variing both temperature and excitation power. 
		{In particular, we show that the $P$ excitons close to the band gap react more sensitively to an increase of the excitation power than states of the associated $D$ exciton multiplet, even though the latter are located at comparatively higher energy. This finding is similar to the one observed when applying an  external electric field, suggesting that the observed behavior arises from internal electric} {fields created by charged impurities that are optically ionized. At laser intensities below 1~$\mu$W/cm$^2$, absorption lines of excitons with $n=29$ are observed. }
	\end{abstract}
	\maketitle
	
	\section{Introduction}

Rydberg excitons in Cu$_2$O are highly-excited electron-hole pairs with large principle quantum numbers $n$. In absorption measurements, excitons with $n=25$ have been observed at a temperature of $T=1.3$~K~\cite{kazimierczukGiantRydbergExcitons2014c}. Later, the Rydberg series was extended at temperatures below 1~K to $n=28$~\cite{heckotterExperimentalLimitationExtending2020}. Employing a  photoluminescence exciation scheme, Versteegh et al. reported signals from excitons with $n=30$~\cite{versteeghGiantRydbergExcitons2021}. 
Excited to high principle quantum numbers $n$, the dipole moment ($\sim n^2$) and polarizability ($\sim n^7$) of Rydberg excitons render them  sensitive to surrounding external electric fields~\cite{heckotterScalingLawsRydberg2017a,gallagher_microwave-optical_2022}. Also, strong interactions with tiny densities of free~\cite{semkatInfluenceElectronholePlasma2019,heckotterRydbergExcitonsPresence2018,stolzScrutinizingDebyePlasma2022} and static charges~\cite{krugerInteractionChargedImpurities2020, bergenLargeScalePurification2023} as well as strong and long-range exciton-exciton interactions~\cite{waltherInteractionsRydbergExcitons2018a, heckotterAsymmetricRydbergBlockade2021} are observed. 

These  interactions give rise to strong optical nonlinearities~\cite{waltherGiantOpticalNonlinearities2018a, zielinska-raczynskaNonlinearOpticalProperties2019, morinSelfKerrEffectYellow2022,  pritchettGiantMicrowaveOptical2024} which can even be enhanced using Rydberg-exciton polaritons~\cite{orfanakisRydbergExcitonPolaritons2022, makhoninNonlinearRydbergExcitonpolaritons2024}. 

Excitons with principal quantum numbers $n$ larger than 15 have been observed solely in high-quality natural crystals while the Rydberg series in artificially grown gemstones is typically limited to $n$ around 15 or lower~\cite{steinhauerRydbergExcitonsCu2O2020, lynchRydbergExcitonsSynthetic2021, orfanakisQuantumConfinedRydberg2021}.  The critical parameter in which natural and artificial samples  differ is still under debate. 

For natural crystals, charged impurities are discussed as a source for electric micro fields that lead to a dissociation of highest $n$ states and limit the Rydberg exciton series~\cite{heckotterExperimentalLimitationExtending2020}.
In that context, the crystal temperature plays a crucial role. At temperatures between 1 and 10~K,  thermal dissociation of shallow impurities may have a stronger impact on  the maximum observable $n$ compared to other thermally driven mechanisms such as  phonon broadening~\cite{heckottertemperature}.

\begin{figure}[ht]
	\centering
	\includegraphics[draft=false,width=0.8\linewidth]{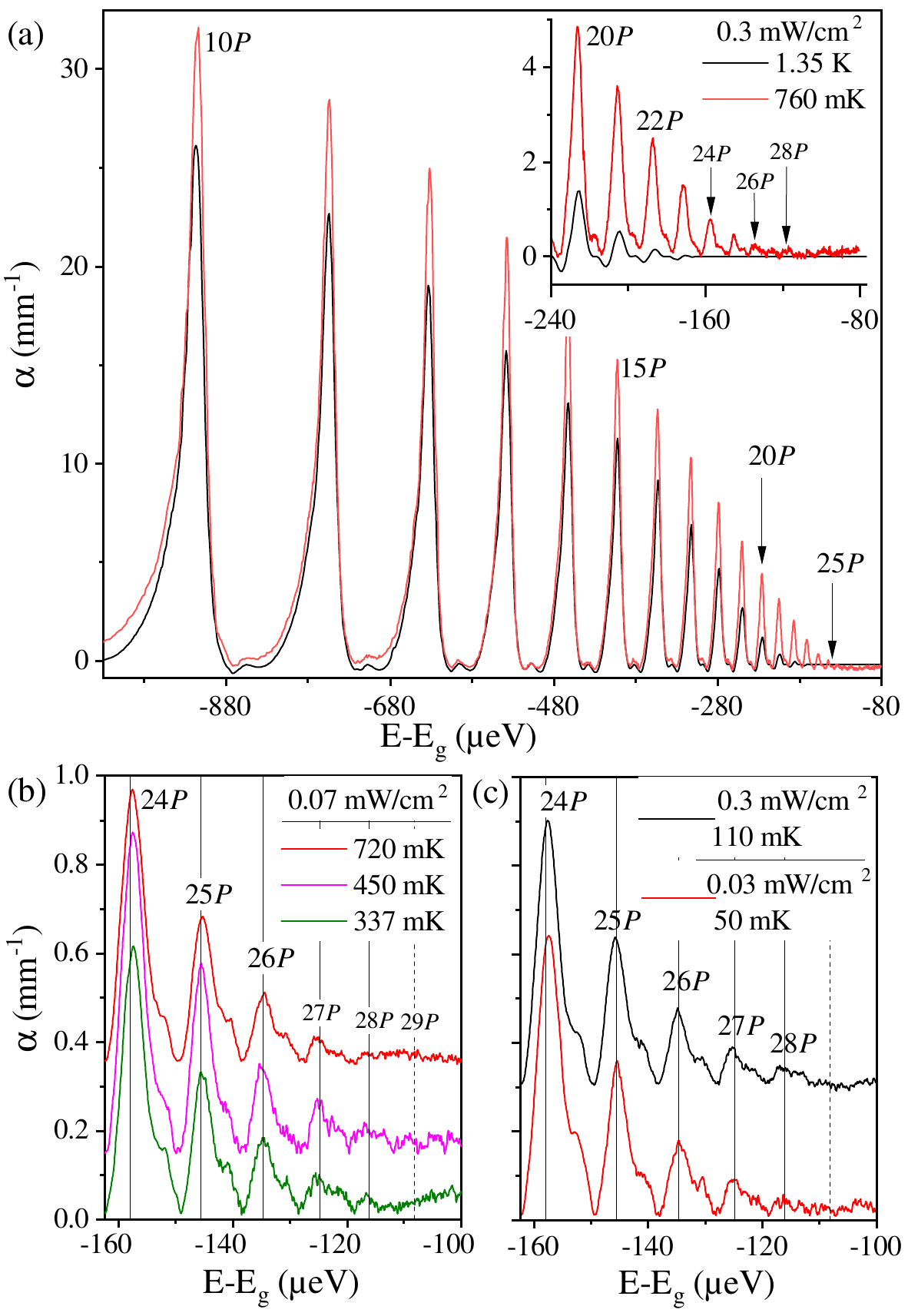}
	\caption{(a) Comparison of absorption spectra measured at 1.35~K and 0.76 ~K with an laser intensity of 0.3~mW/cm$^2$. Below 1~K, the absorption coefficient rises for all $n$. This results in an improvement for the visibility of states up to $n=26$, as highlighted in the inset. (b) Absorption spectra at three different temperatures below 1~K. $n=28$ becomes visible for temperatures equal and below 450~mK. (c) Also at 110~mK and 50~mK $n_{max}=28$ at laser intensities of 0.3~mW/cm$^2$ and 0.03 mW/cm$^2$. The photon energy is given relative to the band gap energy $E_g$. Vertical dashed lines mark the expected resonance energies according to $E_g-Ryd/n^2$, using a Rydberg energy of $Ryd=91$~meV and a band gap energy of $E_g=2.17208$~eV. Adapted with permission from Ref.~\cite{heckotterExperimentalLimitationExtending2020}. } 
	\label{fig:0}
\end{figure}

The absorption spectrum of Rydberg excitons at ultra-low temperatures has been  investigated in Ref.~\cite{heckotterExperimentalLimitationExtending2020}. Figure~\ref{fig:0}(a) shows two exemplary absorption spectra at 1.35~K and 760~mK. 
At a temperature of 760~mK, the peak absorption of exciton lines above $n=20$ certainly increases and states up to $n=26$ become clearly visible. Indications for higher lines exist, but are hard to distinguish from noise. Cooling further down reveals states $n=27$ and $n=28$ clearly from 450~mK downwards (Fig.~\ref{fig:0}(b)). $n_\text{max}$ does not increase further beyond $n=28$ when cooling down to 110~mK or 50~mK (Fig.~\ref{fig:0}(c)). 

Hence, the highest excitable principal quantum number $n_\text{max}$ does not increase much when lowering the temperature below 1~K. At 450~mK as well as 50~mK the highest visible principal quantum number is $n_\text{max}=28$. 
It has been concluded that the density of charged impurities plays an important role for the limitation of $n_\text{max}$. These charged defects generate weak electric stray fields in the material that may dissociate high-$n$ Rydberg excitons~\cite{krugerInteractionChargedImpurities2020}. 
In natural Cu$_2$O gemstones, singly or doubly charged oxygen vacancies ($V_{O^+}$ , $V_{O^{++}}$) are the dominant type of impurities~\cite{koiralaRelaxationLocalizedExcitons2014, frazerVacancyRelaxationCuprous2017, lynchRydbergExcitonsSynthetic2021}. 
The density of impurities may vary across a sample leading to a slight variation of $n_\text{max}$ for different sample positions as well as among different samples. 
With increasing temperature, neutral impurities may undergo thermal dissociation raising the density of charged impurities which lowers the maximum excitable Rydberg state ~\cite{heckottertemperature}. 
Vice versa, at low temperatures thermal dissociation and the density of charged impurities is minimized and more Rydberg states may appear. The highest observable $n$ depends on the remaining density of charged impurities and therefore it depends on the sample quality. 
Moreover, incident laser light can also dissociate neutral impurities by the photo-electric effect~\cite{allen1969photo,kogan1977photoelectric,bergenLargeScalePurification2023}. 
{The spectra in Fig.~\ref{fig:0}(c) were recorded with laser powers of 1~$\mu$W and 100~nW corresponding to intensities of 0.3~mW/cm$^2$ and 0.03 mW/cm$^2$, respectively. Here, the reduction from 1~$\mu$W to 100~nW did not increase the absorption strength of the highest visible state $n=28$~, whereas the observation of Rydberg excitons up to $n=30$ by Versteegh et al.~\cite{versteeghGiantRydbergExcitons2021} was reported at powers as low as 2~nW.}

In order to study the dependence of $n_\text{max}$ on excitation power, we present further absorption spectra of Rydberg excitons in Cu$_2$O at millikelvin temperatures and varying excitation intensities.

\section{Experiment}\label{sec:experiment}
The experimental setup employs a tunable high-resolution dye laser with a spectral resolution of 125~kHz. The sample is a high-quality natural sample that was also used in Ref.~\cite{heckotterExperimentalLimitationExtending2020}. It is placed in a He$^3$-He$^4$ dilution refrigerator (Oxford) to reach temperatures below 1~K. The laser light is focused on the sample and the transmitted intensity is recorded with a photodiode as a function of laser photon energy at different laser intensities and temperatures. 
{For intensities down to 0.08~mW/cm$^2$, a silicon photodiode ('NewFocus 2031') is used. 
To acquire spectra at even lower intensities, the detector is changed to a photodiode 'FWPR-20-SI' by Femto, capable of detecting   powers down to 500~pW. }
The laser power is measured in front of the cryostat 
and translated into an intensity incident on the sample  by including all reflection losses at cryostat windows and the sample surface and dividing by the area of the laser spot at the sample position. The area of the laser spot is 0.3~mm$^2$, except for the measurements shown in Fig.~\ref{fig:3}, where it reads 0.03~mm$^2$.

\section{Variation of laser intensity}\label{sec:laserpower}
\begin{figure}[ht]
	\centering
	\includegraphics[draft=false,width=1\linewidth]{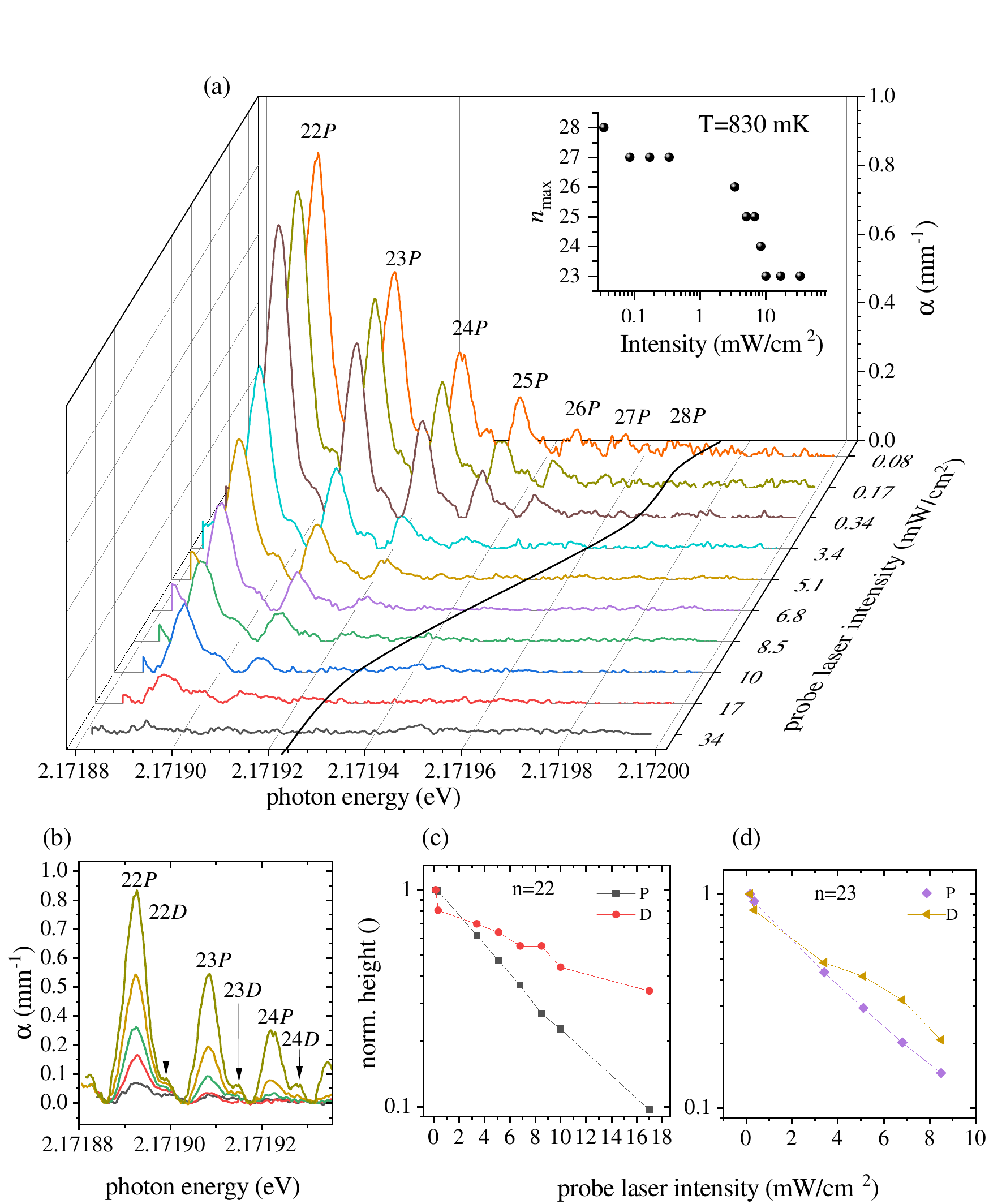}

	\caption{(a)Absorption spectra of Rydberg excitons as a function of laser intensity at a temperature of 830~mK. At an intensity of 0.08~mW/cm$^2$, $n=28$ is visible. The maximum observable $n$ decreases with increasing laser intensity to $n=23$ at 34~mW/cm$^2$. The  inset shows $n_\text{max}$ as a function of intensity.
	(b) On the high-energy side of each $P$ exciton line, small $D$ exciton absorption lines are visible. (c)/(d) Maximum absorption of $P$ and $D$ lines as a function of  laser intensity for (c) $n=22$ and (d) $n=23$. The signal is normalized to unity at lowest applied power. $D$ excitons are less sensitive to the increasing laser intensity compared to $P$ excitons. } 
	\label{fig:1}
\end{figure}

We start with a variation of laser intensity at a temperature of 830~mK, as shown in Figure~\ref{fig:1} in a waterfall diagram.  
The spectra show the absorption lines of exciton states from $n=22$ onwards. The continuous phonon background and continuum absorption are subtracted. At this temperature, $n_\text{max}$ depends strongly on the laser intensity. At the lowest intensity of 0.08~mW/cm$^2$ the spectrum reveals lines up to $n=28$.  With increasing intensity $n_\text{max}$ decreases down to $n=23$ at 34~mW/cm$^2$. 
The solid line indicates the end of the Rydberg series as a guide to the eye. This energy, at which the Rydberg series ends,   is referred to as the shifted band gap. 
At 830~mK, thermal dissociation is low and most charged impurities are in the neutral state. 
Instead, the incident laser light may induce photo-electric dissociation of neutral impurities and a rise of the density of charged impurities with increasing laser intensity. 
Also, the  laser may excite a low-density electron-hole plasma that leads to a lowering of the band gap and attenuation of states close to the band gap. 
Both effects scale equally with the fourth root of the density of either impurities~\cite{krugerInteractionChargedImpurities2020} or electron-hole pairs~\cite{semkatInfluenceElectronholePlasma2019}. 
In both cases, the highest Rydberg states vanish with increasing laser intensity. 

We note that $n=28$ is observed {here} at a temperature of 830~mK already with 0.08~mW/cm$^2$ incident intensity, while in Fig.~\ref{fig:0}(b) a distinct feature for $n=28$ appears only from 450~mK downwards at a comparably low intensity of 0.07~mW/cm$^2$. This highlights the role of the particular measured spot on the sample: {The plasma density induced by the  laser is comparable in both cases since the incident power is almost equal. Hence, the main difference is a varying sample position and the number of impurities that is ionized by the laser presents the main limiting parameter for the observation of higher $n$.} The density of impurities may vary among the sample, even at lowest temperatures, thus leading to a varying quality of the spectra among the same sample. 

{A closer look at the series of absorption lines reveals a shoulder on the high energy side of each $P$ exciton line, see Fig.~\ref{fig:1}(b). 
These additional absorption features have been attributed to even parity $D$ excitons that become optically allowed by the Stark effect induced by the electric fields of charged impurities~\cite{heckotterExperimentalLimitationExtending2020,heckotter_neutralization_2023}. 
{A characteristic feature is that the energy splitting between the $P$ and $D$ states is almost independent of the involved principal quantum number, while one would expect the width of the fine structure multiplet to decrease with $n$~\cite{gallagherRydbergAtoms1994}. This finding is another indication that electric fields are affecting the exciton lines. When applying an electric field, the $P$ exciton energy becomes lowered, while the $D$ exciton energy increases, reflecting the evolution to a Stark fan. This explains the constant splitting. The electric field influence is also manifested by the relative absorption strength of the two peaks. While for $n=24$ the $D$ exciton is still a weak high energy shoulder of the $P$ exciton, it becomes more and more prominent relative to the $P$ exciton with increasing principal quantum number. For $n=27$ and 28 both peaks have almost the same amount of absorption.}
These $D$ excitons appear less sensitive to the increasing laser intensity compared to the adjacent $P$ excitons. This is shown in Fig.~\ref{fig:1}(c) and (d), where the  absorption of $P$ and $D$ excitons for $n=22$ and $=23$ is shown as a function of  laser intensity.  
{For $n=22$, the peak absorption of the $P$ exciton drops much more strongly with laser power than the $D$ exciton. Both states are {still} visible at the highest powers, as the effective bandgap has not yet crossed their binding energies. In this process, the absolute  difference in peak absorption has dropped from almost one order of magnitude to a factor of two. Also for $n=23$, a similar behavior is found: the $P$ exciton absorption maximum drops faster then that of the $D$ state, which is consistent with the scaling for even higher principal quantum numbers. }
This observation is in line 
with the behavior of high-$L$ excitons in  electric fields~\cite{heckotterDissociationExcitonsMathrmCu2018a}: }
Excitons with different $L$ and opposite parity become mixed by the electric field leading to the Stark effect. 
	In a Stark fan of a multiplet with a particular $n$, states on the high-energy side of the Stark fan have wavefunctions with a dipole moment aligned opposite to the electric field and a high probability density that is shifted away from the tunnel barrier. 
	This  renders states on the high energy side of a Stark fan less sensitive to dissociation compared to states on the low energy side. 

\begin{figure}[ht]
	\centering
	\includegraphics[draft=false,width=\linewidth]{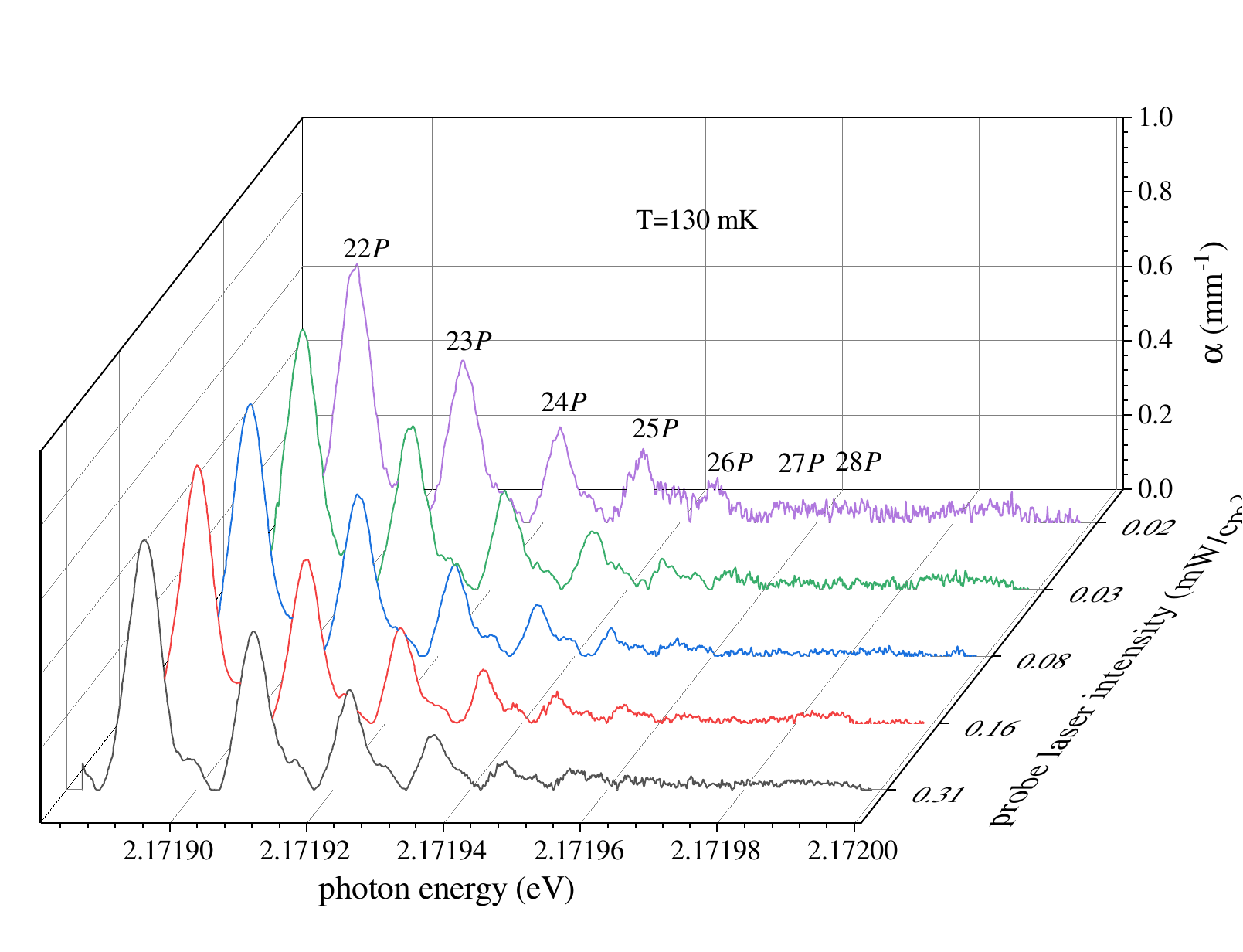}
	
	\caption{Absorption spectra of Rydberg excitons for series of laser intensities below 0.3~mW/cm$^2$ at a temperature of 130~mK. The vertical dashed lines mark the expected resonance energies according to $E_g-Ryd/n^2$, using a Rydberg energy of $Ryd=91$~meV and a bandgap energy of $E_g=2.17208$~eV.}
	\label{fig:3}
\end{figure}

{Coming back to the maximum observable $n_\text{max}$, we conclude that low intensities increase the number of Rydberg states at a particular sample position: 
In order to observe Rydberg exciton states with $n>25$, intensities below 5~mW/cm$^2$ and temperatures below 1~K are required. }

Hence, absorption spectra were measured at even lower  laser intensities.  Also, the temperature was further reduced to minimize  thermal dissociation.  Figure~\ref{fig:3} shows the absorption of exciton states from $n=22$ onwards for laser intensities from 0.31~mW/cm$^2$ to 0.02~mW/cm$^2$ at 130~mK. We note that the sample position differs from the one in Fig.~\ref{fig:1}. In order to improve the signal to noise ratio, five scans are taken and averaged for each laser intensity. 
At 0.31~mW/cm$^2$, a weak absorption feature for $n=27$ is clearly distinguishable from noise and determines the highest Rydberg state.  
When reducing the intensity even further, no higher lines emerge. 
Interestingly, at 0.08~mW/cm$^2$, $n=28$ is not visible, although it is visible in Fig.~\ref{fig:1} at the same intensity and an even higher temperature. This underlines the role of the position of measurement spot on the sample for $n_\text{max}$, which finally determines  $n_\text{max}$  via the impurity density and dominates over the  temperature. 
At lowest intensity of 0.02~mW/cm$^2$, the noise level overcomes any possible additional exciton feature. 

\begin{figure}[ht]
	\centering
	\includegraphics[draft=false,width=0.8\linewidth]{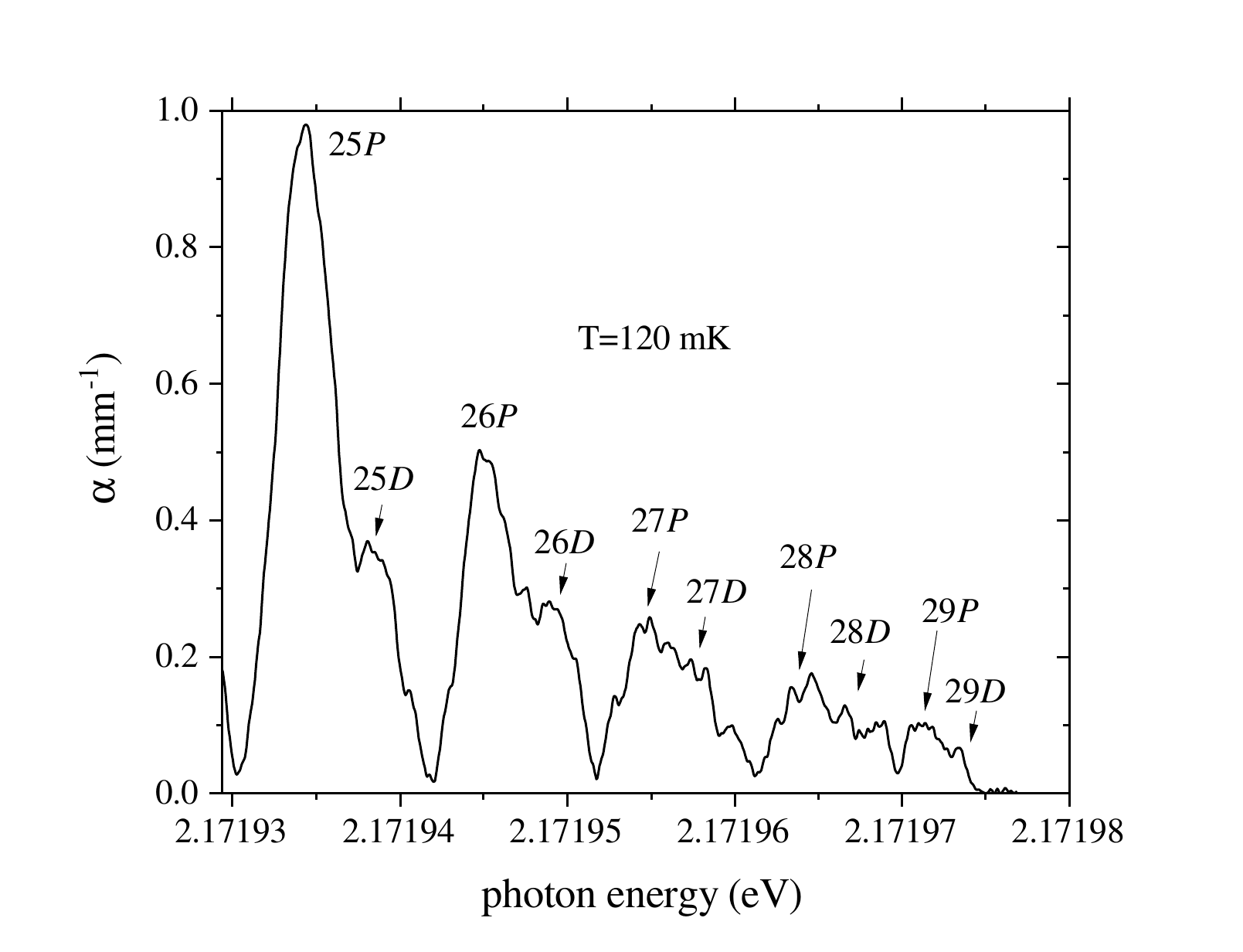}
	
	\caption{Absorption spectrum at 120~mK and a laser intensity of 0.0003~mW/cm$^2$. The spectrum is an average of five single absorption spectra to reduce the noise.}
	\label{fig:4}
\end{figure}

The lowest intensity where a signal was detectable amounts to $3\times 10^{-4}$~mW/cm$^2$. The resulting averaged spectrum at a temperature of 120~mK is shown in Fig.~\ref{fig:4}. Also here, the continuous background is subtracted.  At this low excitation intensity, an indication of $n=29$ is observable before the spectrum becomes flat.

\section{Summary and discussion}\label{sec:Conclusions}
We extended the experimental parameter space for absorption spectra of Rydberg excitons in Cu$_2$O by varying the laser intensity at different temperatures below 1~K. 
We do not find a clear dependence of $n_\text{max}$ on temperature, as long as it stays below 1~K. 
$n_\text{max}$ rather depends on the sample quality at the particular measured spot on the sample. 
At a particular good spot on the sample, we observe $n_\text{max}=28$ already at a temperature of 830~mK and low intensity. 
Moreover, at temperatures below 1~K, the maximum observable $n$ depends strongly on the applied laser intensity. 
To observe Rydberg excitons with principal quantum numbers higher than $n=25$, the laser intensity has to be reduced below 5~mW/cm$^2$. 
At 0.0003~mW/cm$^2$, an indication of $n=29$ becomes apparent. 
We conclude that at temperatures below 1~K, laser intensity is the crucial experimental parameter to observe high-$n$ Rydberg states, given that the best sample position is probed. This is in agreement with Ref.~\cite{versteeghGiantRydbergExcitons2021}, where $n=27$ was observed at a power of 10~nW and $n=30$ only at 2~nW. 
The laser field photo-ionizes impurities and creates electric microfields in the sample. This is supported by the behavior of $D$ exciton absorption features that are more stable against increasing laser power than $P$ excitons.

\section*{Acknowledgments}
We acknowledge the financial support by the Deutsche Forschungsgemeinschaft through projects 504522424 and 427377618. 

\clearpage
\input{mKfollowup.bbl}
\end{document}

%% file: mKfollowup.bbl
%